\documentclass[groupedaddress,
  aps,
  pra,
  amsmath,amssymb,
  reprint,
  longbibliography,
]{revtex4-1}
\usepackage{graphicx,color}
\usepackage{amsmath}
\usepackage{natbib}
\usepackage{epsfig}
\begin{document}

\title{Note: Sound velocities of generalized Lennard-Jones ($n-6$) fluids near freezing}

\author{Sergey A. Khrapak}
\affiliation{Bauman Moscow State Technical University, 105005 Moscow, Russia;\\ Joint Institute for High Temperatures, Russian Academy of Sciences, 125412 Moscow, Russia}


\date{\today}

\begin{abstract}
In a recent paper [S. Khrapak, Molecules {\bf 25}, 3498 (2000)] the longitudinal and transverse sound velocities of conventional Lennard-Jones systems at the liquid-solid coexistence were calculated. It was shown that the sound velocities remain almost invariant along the liquid-solid coexistence boundary lines and that their magnitudes are comparable with those of repulsive soft sphere and hard sphere models at the fluid-solid phase transition. This implies that attraction does not affect the magnitude of sound velocities at the fluid-solid phase transition. This paper provides further evidence to this by examining the generalized Lennard-Jones $n$-6 fluids with $n$ ranging from $12$ to $7$ and demonstrating that the steepness of repulsive term has only a minor effect on the magnitude of the sound velocities.      
\end{abstract}

\maketitle

\section{Introduction}

Sound velocities are very important characteristics of a substance. They are directly related to long-wavelength excitations -- phonons, which play a crucial role in condensed matter, materials science, and soft matter. Therefore it is very important to understand mechanisms that can affect and regulate sound velocities in various situations.

In simple atomic systems with smooth pairwise interaction potentials the sound velocities can be expressed as sums over atoms involving the first and second derivatives of the interaction potential (see below). In isotropic situations (in gases and liquids) these sums can be also expressed as integrals involving the radial distribution function $g(r)$. The sound velocities are thus completely determined by the shape of the interaction potential and the atomic structure of the system (which are interrelated).       
 
There have been indications that the shape of the interatomic interaction potential may not be a very important factor affecting the sound velocities in special regions of the phase diagram. For example, the ratio of the sound to thermal velocity of many liquid metals and metalloids has about the same value $\simeq 10$ at the melting temperature.~\cite{IidaBook,BlairsPCL_2007,RosenfeldJPCM_1999} A close value ($\simeq 9.5$) was reported from experiments with solid argon at the melting temperature.~\cite{IshizakiJCP1975} Values close to $\simeq 9$ were reported for solid hydrogen and deuterium along the melting curve.~\cite{LiebenbergPRB1978} Rosenfeld pointed out that this ``quasi-universal'' property is also shared by the hard-sphere (HS) model.~\cite{RosenfeldJPCM_1999} More recently, it was demonstrated that this property is also exhibited by the purely repulsive soft inverse-power-law (IPL) model in a wide range of IPL exponents.~\cite{KhrapakJCP2016}

A generalization of the celebrated Lindemann's criterion of melting of classical solids was proposed recently on the basis of statistical mechanics arguments.~\cite{KhrapakPRR2019} With this generalized formulation the expressions for the melting temperature are equivalent in three dimensions (3D) and two-dimensions (2D). Moreover, independently of dimensionality, the melting condition predicts that the ratio of the transverse sound velocity to the thermal velocity reaches a quasi-universal value along the melting curve (the magnitudes can be different in 3D and 2D)~\cite{KhrapakPRR2019,KhrapakJCP2018}.  

Motivated by these observation, it was recently investigated in detail how the longitudinal and transverse sound velocities behave at the liquid-solid phase transition of 3D Lennard-Jones (LJ) systems~\cite{KhrapakMolecules2020}. It was observed that the sound velocities keep quasi-universal values along the liquid-solid coexistence boundaries and that their magnitudes are comparable with those of repulsive soft sphere and hard sphere models. Hence, it was concluded that attraction does not affect the magnitude of sound velocities at the fluid-solid phase transition. This paper provides further evidence to this conclusion by examining the generalized Lennard-Jones $n$-6 fluids with $n$ ranging from $12$ to $7$ to vary the relative strength of attraction. It will be documented that the steepness of repulsive term has relatively weak effect on the magnitudes of longitudinal and transverse sound velocities.       

\section{Materials and Methods}
%

In this work the generalized $n$-6 LJ potential is considered:
\begin{equation}\label{LJ}
\phi(r)= C_n\epsilon\left[(\sigma/r)^n - (\sigma/r)^6\right],  
\end{equation}
where
\begin{displaymath}
C_n=\left(\frac{n}{n-6}\right)\left(\frac{n}{6}\right)^{\frac{6}{n-6}}.
\end{displaymath}
Here $\epsilon$ and $\sigma$ are the energy and length scales and the exponent $n$ determines the steepness of the repulsive term in the potential (\ref{LJ}). The reduced pressure, energy, density, and temperature are expressed as  $P_*= P\sigma^3/\epsilon$, $u_*=U/N\epsilon$, $\rho_*=\rho\sigma^3$, and $T_*=T/\epsilon$, respectively.

For a spherically symmetric pairwise interaction potential $\phi(r)$, the longitudinal and transverse velocities can be expressed (in 3D) as follows~\cite{Schofield1966,BalucaniBook,Takeno1971}
\begin{eqnarray}\label{long}
mc_l^2 &=& \frac{1}{30}\left\langle\sum_j\left[2r_j\phi'(r_j)+3r_j^2\phi''(r_j)\right] \right\rangle,  \\ \label{trans}
mc_t^2 &=& \frac{1}{30}\left\langle \sum_j\left[4r_j\phi'(r_j)+r_j^2\phi''(r_j)\right]\right \rangle.
\end{eqnarray}   
Here $c_{l}$ and $c_t$ are longitudinal and transverse elastic sound velocities, $m$ is the particle mass,  the sums run over all neighbours of a test particle, and primes denote derivatives of the interaction potential with respect to the distance $r$. The averaging $\langle ... \rangle$ denotes that we perform summation over all $N$ test particles in the system and then divide the sum by $N$. In isotropic fluids such an averaging is often replaced an integral involving the radial distribution function,~\cite{ZwanzigJCP1965,Schofield1966,Hubbard1969} $\langle\sum_j(...)\rangle \rightarrow 4\pi\rho \int(...)r^2g(r)dr$, where $g(r)$ is the radial distribution function. For further details regarding the evaluation of sound velocities and their relation to the instantaneous elastic moduli see Ref.~\cite{KhrapakMolecules2020}. 

For a pairwise interaction potential the excess (over the ideal gas) contributions to the energy, $u_{\rm ex}$, and pressure, $p_{\rm ex}$, can be expressed via summations similar to those used above (we omit averaging to keep the notation compact):
\begin{equation}\label{sums}
u_{\rm ex}=\frac{1}{2T}\sum_j\phi(r_j), \quad p_{\rm ex} = -\frac{1}{6T}\sum_jr_j\phi'(r_j),
\end{equation} 
where the reduced units are used: $u_{\rm ex}=U_{\rm ex}/NT$, $p_{\rm ex}=P_{\rm ex}/\rho T$ (the temperature $T$ is measured in energy units, $k_{\rm B}=1)$. It is obvious then that we can express the reduced sound velocities in terms of reduced excess energy and pressure. After simple straightforward algebra we get 
\begin{eqnarray}\label{c_l}
c_l^2/v_{T}^2 &=& {\mathcal A}_l(n) u_{\rm ex}+{\mathcal B}_l(n) p_{\rm ex}, \\
c_t^2/v_{T}^2 &=& {\mathcal A}_t(n) u_{\rm ex}+{\mathcal B}_t(n) p_{\rm ex},\label{c_t}
\end{eqnarray}
with the $n$-dependent numerical coefficients
\begin{eqnarray}
{\mathcal A}_l(n) &=&-\frac{6n}{5}, \quad {\mathcal B}_l(n) = \frac{n(3n+1)-114}{5(n-6)}, \\
{\mathcal A}_t(n) &=& -\frac{2n}{5}, \quad {\mathcal B}_t(n) = \frac{n(n-3)-18}{5(n-6)}.\label{t}
\end{eqnarray}  
The sound velocities in Eqs.~(\ref{c_l}) and (\ref{c_t}) are expressed in units of thermal velocity, $v_{\rm T}=\sqrt{T/m}$. Equations (\ref{c_l}) - (\ref{t}) can be considered as generalization of the Zwanzig and Mountain result for the conventional (12-6) LJ potential. In this special case the constants are ${\mathcal A}_l(12) = -72/5$, ${\mathcal B}_l(12) = 11$, ${\mathcal A}_t(12) = -24/5$, and ${\mathcal B}_t(12) = 3$~\cite{ZwanzigJCP1965,KhrapakMolecules2020}.  

The translation to LJ units is straightforward by recognizing that
\begin{eqnarray}
P_* \equiv \frac{P\sigma^3}{\epsilon}=\frac{\rho T(1+p_{\rm ex})\sigma^3}{\epsilon}=\rho_*T_*(1+p_{\rm ex}), \\
u_*\equiv \frac{U}{N\epsilon}=\frac{NT(3/2+u_{\rm ex})}{N\epsilon}=T_*(3/2+u_{\rm ex}).
\end{eqnarray} 

\section{Results and discussion}
  
\begin{figure}
\includegraphics[width=8cm]{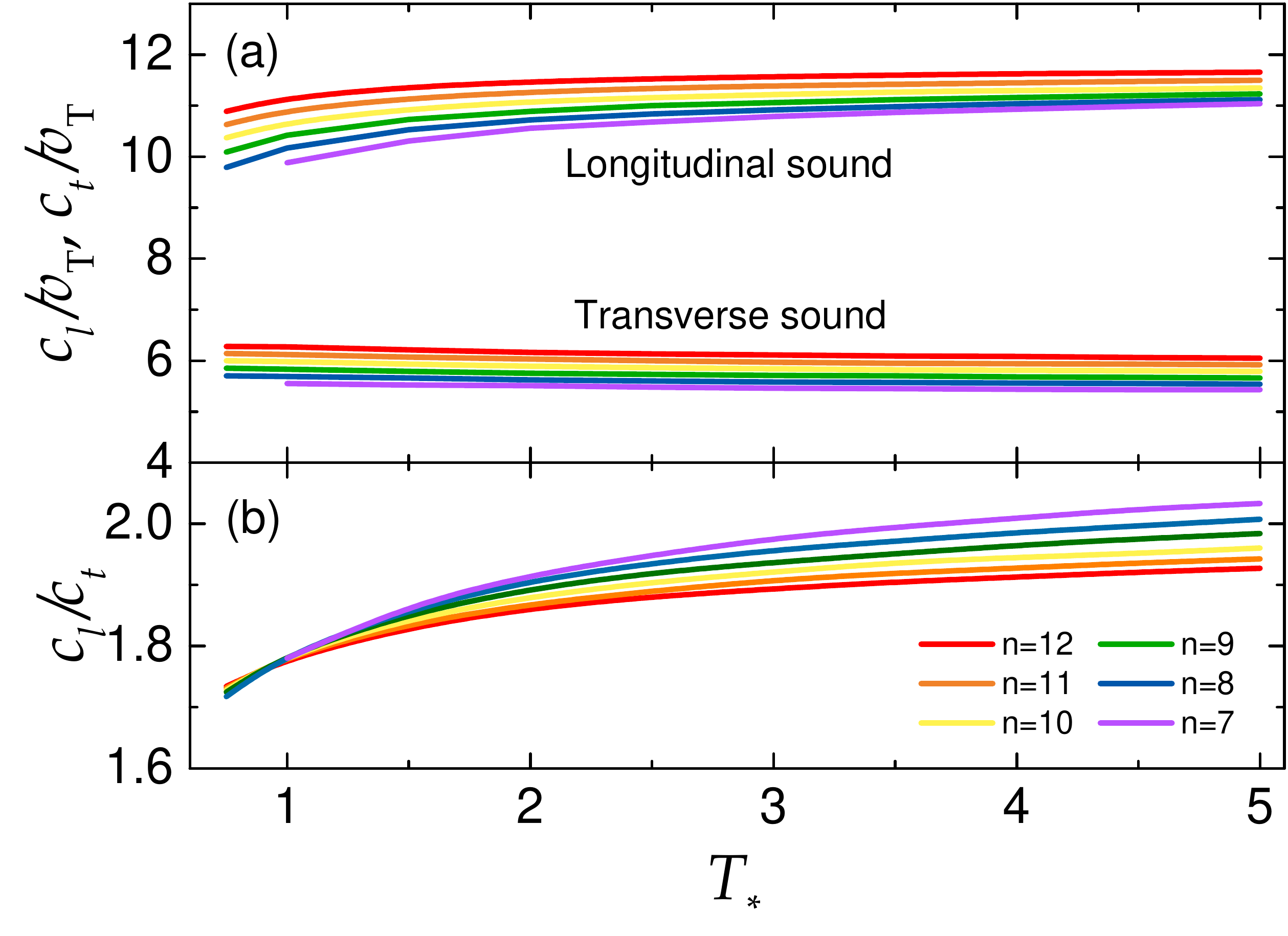}
\caption{Longitudinal ($c_l$) and transverse ($c_t$) sound velocities of generalized LJ $n$-6 fluids at freezing vs the reduced temperature $T_*$. In (a) sound velocities are expressed in units of the thermal velocity $v_{\rm T}$; Upper curves are for the longitudinal mode, lower curves are for the transverse mode. In (b) the ratio of the longitudinal-to-transverse sound velocities, $c_l/c_t$is plotted.}
\label{Fig1}
\end{figure}

The sound velocities of generalized ($n$-6) LJ fluids at freezing condition have been evaluated using the freezing parameters ($T_*$, $P_*$, and $\rho_*$) tabulated by Sousa {\it et al}.~\cite{SousaJCP2012} (simple analytical fits for the freezing and melting curves are also available~\cite{KhrapakAIPAdv2016}, but were not used here). Note that the reduced energy tabulated there correspond to the potential contribution only, $u_*=T_*u_{\rm ex}$. The results are presented in Figure~\ref{Fig1}.    


In Figure~\ref{Fig1}(a) the reduced sound velocities are
plotted versus the reduced temperature. The main observations are as follows. For each repulsive exponent $n$ the longitudinal and the transverse sound velocities remain practically constant along the freezing curve. The approximate constancy of the transverse sound velocity at the fluid-solid phase transition can be interpreted employing the Lindemann's melting criterion arguments~\cite{Lindemann}.
In one of the formulations of this famous criterion, the melting condition can be expressed in terms of the infinite frequency (instantaneous) elastic longitudinal and shear moduli and thus in terms of the longitudinal and transverse sound velocities~\cite{BuchenauPRE2014,KhrapakPRR2019}. In particular, the condition contains a sum over inverse longitudinal and shear moduli, each multiplied by the number of the corresponding collective modes. Since the shear modulus is smaller than the longitudinal one and there are two shear and one longitudinal mode in three-dimensional systems, the contribution from shear modes dominates (the effect known as shear dominance) thus leading to the conclusion that $c_t/v_T$ should be approximately constant at the fluid-solid phase transition~\cite{KhrapakPRR2019}.  
The approximate constancy of the longitudinal sound velocity can be explained as follows. Although the effective softness of a given LJ ($n$-6) potential somewhat varies along the freezing curve~\cite{KhrapakJCP2011_3} (i.e. with $T_*$), the variation of $c_l/v_T$ with effective softness is rather weak (provided the potential is not too soft and long-ranged), see Fig. 1 from Ref.~\cite{KhrapakMolecules2020}.         

Regarding the dependence of the reduced sound velocities on the repulsive exponent $n$ at a fixed $T_*$, a very weak, but systematic tendency of increasing $c_l/v_{\rm T}$ and $c_t/v_{\rm T}$ with $n$ is observed. This tendency is consistent with a very weak increase of the reduced sound velocities when the HS interaction limit is approached, which has been documented in Fig. 1 from Ref.~\cite{KhrapakMolecules2020}. Increasing $n$ leads to somewhat steeper repulsion and the reduced sound velocities increase accordingly, although very weakly. 

To a reasonable approximation, the reduced sound velocities can be approximated by constant values of  $c_l/v_{\rm T}\simeq 11$ and $c_t/v_{\rm T}\simeq 6$. Deviations from accurate values are within $\sim 10\%$ for all exponents $n$ in the range of $T_*$ considered. The sound velocity of purely repulsive soft inverse-power-law interaction potentials fall in the same range. For instance, the sound velocities of IPL-12 system at melting is $c_l/v_{\rm T}\simeq 11.7$ and $c_t/v_{\rm T}\simeq 5.8$~\cite{KhrapakMolecules2020}. In this sense it is legitimate to say that the presence of long range attraction has a rather minor effect on the sound velocities of simple fluids near the fluid-solid phase transition.  

The ratio of the longitudinal-to-transverse sound velocities is plotted in Fig.~\ref{Fig1}(b). It exhibits a slow monotonous increase with $T_*$. In addition, at higher temperatures there is an observable decline in $c_l/c_t$ with increasing $n$. This tendency should be expected. For example, for soft Yukawa (screened Coulomb) interactions potential the ratio $c_l/c_t$ increases when the interaction becomes softer (screening length increases) and diverges on approaching the one-component plasma limit (see e.g. Fig. 2 from Ref.~\cite{KhrapakPoP2019}). At the same time for all $n$ considered here and the range of $T_*$ studied the ratio $c_l/c_t$ remains in a relatively narrow range, between $\simeq 1.7$ and $\simeq 2.0$.    

\section{Conclusion}

The longitudinal and transverse sound velocities of the generalized ($n$-6) Lenard-Jones fluids have been evaluated at freezing conditions for integer $n$ between $7$ and $12$. For this purpose, the relation between the sound velocities (or elastic moduli) and excess energy and pressure derived by Zwanzig and Mountain~\cite{ZwanzigJCP1965} for the conventional (12-6) LJ potential has been generalized to the case of ($n$-6) potential. The required freezing densities, pressures, and internal energies for temperatures in the range from $T_*=1$ to $T_*=5$ have been taken from Ref.~\cite{SousaJCP2012}.    

The calculated ratios of sound velocities to the thermal velocity are practically constant along the freezing curves with the typical values $c_l/v_{\rm T}\simeq 11$ and $c_t/v_{\rm T}\simeq 6$. They are comparable with reduced sound velocities of repulsive soft sphere  (IPL) and HS fluids at freezing. Thus, roughly speaking neither the presence of the long-range attraction nor the variation of the exponent in the repulsive term do not lead to considerable variations in the sound velocities. At the same time, minor systematic trends have been identified and discussed. This trends can be convincingly explained on the basis of the known (weak) dependence of the reduced sound velocities at freezing on the effective softness of the interaction potential.




\bibliography{Additivity}

\end{document}